\documentclass[pra,twocolumn]{revtex4}
\usepackage{graphicx}
\usepackage{amssymb, amsmath}
\usepackage{subfigure}

\begin{document}
\title{Superradiance and phase multistability in circuit quantum electrodynamics}
\author{M. Delanty}
\email{michael.delanty@mq.edu.au}
\author{S. Rebi\'c and J. Twamley}
\address{Centre of Excellence for Engineered Quantum Systems (EQuS), Department of Physics and Astronomy, Macquarie University, Sydney, NSW 2109, Australia}
\begin{abstract}
By modeling the coupling of multiple superconducting qubits to a single cavity in the circuit-quantum electrodynamics (QED) framework we find that it should be possible to observe superradiance and phase multistability using currently available technology. Due to the exceptionally large couplings present in circuit-QED we predict that superradiant microwave pulses should be observable with only a very small number of qubits (just three or four), in the presence of energy relaxation and non-uniform qubit-field coupling strengths. This paves the way for circuit-QED implementations of superradiant state readout and decoherence free subspace state encoding in subradiant states. The system considered here also exhibits phase multistability when driven with large field amplitudes, and this effect may have applications for collective qubit readout and for quantum feedback protocols.
\end{abstract}


\maketitle

\section{Introduction}

In 1954, Dicke \cite{Dicke}  noted that the intensity of radiation from a gas could be enhanced by confining it in a region smaller than the wavelength of emitted radiation, a phenomenon called superradiance. The collective emission intensity, which scales with $\propto N^2$,  is of fundamental interest in quantum optics and has been the subject of many theoretical and experimental works \cite{Agarwal1974, HarocheReview, BrandesReview}.  With greater control over a range of quantum systems, it is now possible to observe superradiance in quantum dots \cite{SR_QDs}, Bose Einstein condensates \cite{SRBEC2003} and nitrogen vacancy centers in diamond \cite{NVSper}. Furthermore superradiance and the  related phenomena of subradiance have quantum information based applications such as  decoherence free subspace state encoding in subradiant states \cite{KnightDFS} and superradiant state readout \cite{Briggs}.

Superradiance has been difficult to observe experimentally as decoherence and dissipation destroy the required build up of correlations within the ensemble \cite{Agarwal1974, HarocheReview, BrandesReview}. Plagued by these losses, experiments have required large ensembles to observe the $ \propto N^2$ intensity characteristic of superradiance \cite{ SR_QDs, SRBEC2003, NVSper, Clearest, Lamb, HarocheSR}. This has inhibited any detailed study of emission dynamics or measurement of correlations within the ensemble, leaving the phenomenon demonstrated in principle yet not explored in detail.

Another phenomenon that is difficult to observe experimentally is phase bistability. In 1991, Alsing and Carmichael, and independently Kilin and Krinitskaya, showed that a strongly driven qubit-cavity system can undergo a symmetry breaking phase transition to a state where the intracavity field is displaced in phase, dependent on the qubit's state \cite{KillinSingle, Alsing}. This state has the remarkable property that the qubit-cavity system can be described as  two coupled, driven and damped harmonic oscillators. Kilin and Krinitskaya also predicted that a similar phenomenon, phase multistability, occurs when many qubits couple to a single cavity mode \cite{KilinMulti}. Phase multistability describes the phenomenon, where a coherently driven multi-qubit cavity system displaces the cavity field in phase depending on the collective state of the qubits.  Phase multistability for $N>1$ has yet to be observed.

\begin{figure}
\centering
{\includegraphics[scale=0.17]{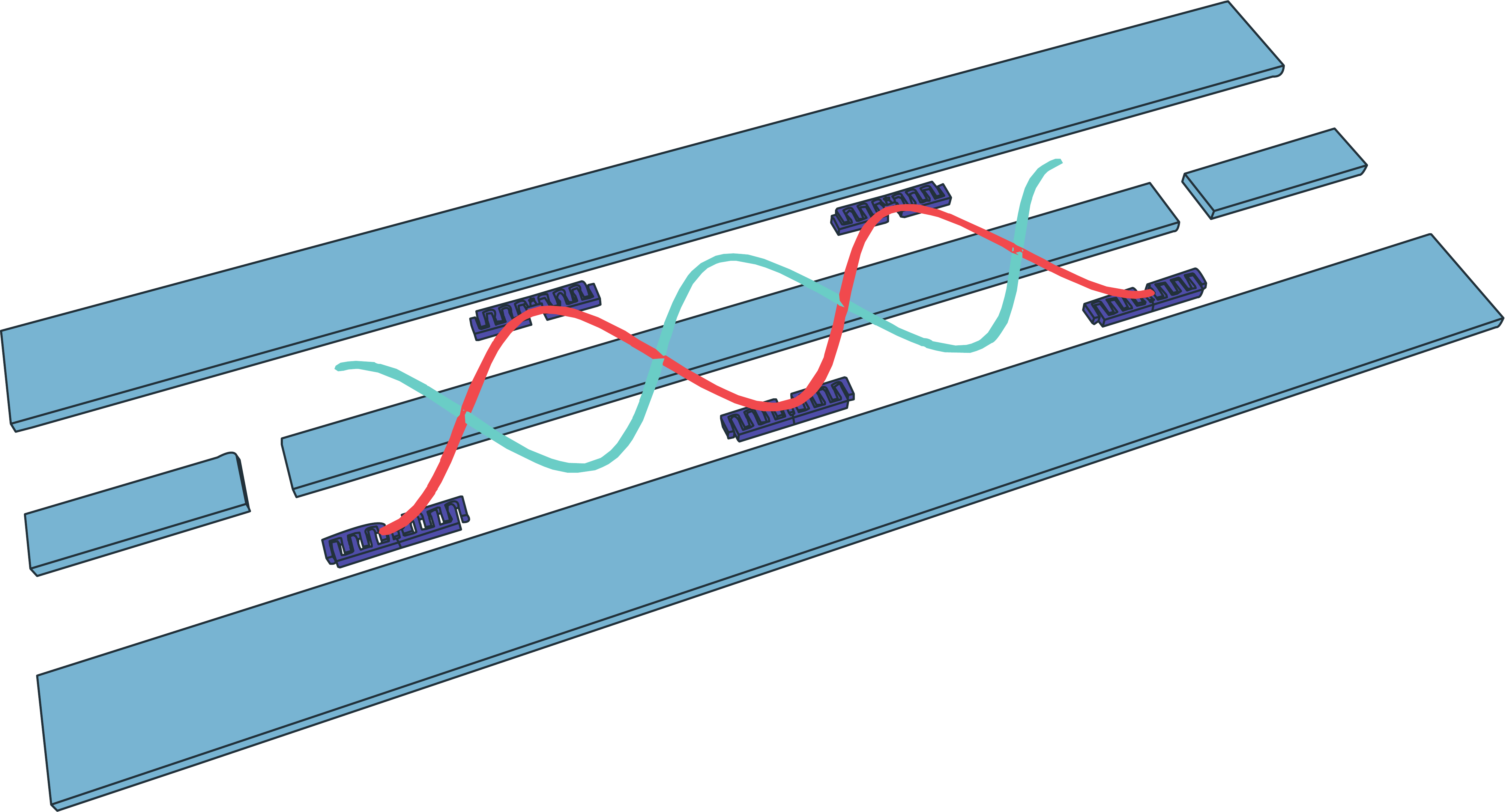}}
\caption{ Five transmon qubits (dark blue) coupled to the quantized field of a transmission line resonator (sinusoids).
 }
 \label{ManyTransmonimage}
\end{figure}

Recently, several promising experiments in circuit-QED have demonstrated various quantum information based tasks such as three qubit entanglement, multi-qubit measurement and the realisation of simple quantum algorithms \cite{Dicarlo, Dicarlo3Qubit, MA}. With such a large degree of control over system dynamics the question naturally arises, ``Can we use superconducting circuits to probe many body collective effects that were previously inaccessible to experiment?". By demonstrating that small sample superradiance and phase multistability can be observed in circuit-QED, we provide an affirmative answer to this question.   Circuit-QED is an ideal platform to observe small ensemble superradiance  and phase multistability due to small qubit losses and large qubit-field coupling rates  \cite{Fink}.

This article is organized as follows. In section \ref{SR}, we show that circuit-QED can probe the small ensemble regime of superradiance, allowing its detailed investigation.  In section \ref{PM} we extend the system by adding a coherent driving of the resonator mode. This leads to the phenomenon of phase multistability which is analysed in detail. Our conclusions are presented in section \ref{con}.

\section{Superradiance \label{SR}}

Here, we study several transmon qubits coupled to the field of a transmission line resonator (TLR) (Fig.~\ref{ManyTransmonimage}). In the frame rotating at angular frequency $\omega$ the dynamics of the system are described by the master equation,
\begin{eqnarray} \label{ME_NCPB}
 \dot{\rho} &=& -i [ H, \rho] +\frac{\kappa}{2}  \mathcal{D}[a] \rho + \sum^N_{j=1} \frac{\gamma^s_j}{2} \mathcal{D}[ \sigma_j^-] \rho + \frac{\gamma^p_j}{2}  \mathcal{D} [\sigma_j^z]  \rho
\end{eqnarray}
where $\gamma^s_j$ and $ \gamma^p_j$ are  for the $j$th qubit the energy relaxation and dephasing rate respectively, $\kappa$ is the resonator decay rate, $\mathcal{D}[ A] \rho = 2 A \rho A^{\dagger} -A^{\dagger} A \rho - \rho A^{\dagger}A $ and $\hbar = 1$. The qubit-resonator system evolves in the interaction picture  under the Tavis-Cummings Hamiltonian \cite{TavisCummings},
\begin{equation} \label{HNCPBcollect}
H = \Delta_r a^{\dagger} a + \sum^N_{j=1}\left( \frac{\Delta_{q, j} }{2} \sigma_j^z + g_{N,j} ( \sigma^-_j a^{\dagger} + a \sigma^+_j) \right),
\end{equation}
where, the $j$th qubit couples to the field at the rate   $ g_{N,j} $,  $\Delta_{q, j}=\omega_{q,j}-\omega$ is the detuning from the $j$th qubit transition frequency $\omega_{q, j}$ and  $\Delta_r=\omega_r-\omega$ is the detuning from the resonator frequency $\omega_r$. The system is initially taken to be in the state where the qubits are prepared in a symmetrically excited state and there are no photons in the cavity, $|\psi (0)\rangle = |0  \rangle \bigotimes |e, e, \dots, e \rangle$. Due to the Tavis-Cummings interaction, energy is transfered to the resonator. We probe the superradiant emission of these photons from the TLR in the ``bad cavity" limit~\cite{MandelBook}, where the photons escape the cavity before re-absorption by the qubits.

In order to obtain an analytic expression for the intensity of emitted photons from the TLR, it is necessary to make several assumptions. First, we assume that each qubit  couples identically to the field mode ($  g_{N,j}  \approx \bar{g}_N$), as non-uniform coupling rates induce subradiant transitions. This can be achieved by optimizing the placement of qubits at field antinodes \cite{Dicarlo, Fink}. Secondly, in the bad cavity limit it is assumed that  the field in the resonator decays at a faster rate than the average qubit-field coupling rate, $\kappa \gg \bar{g}_N$. Lastly, it is assumed that the dissipative processes, dephasing ($\gamma_j^p$) and energy relaxation ($\gamma_j^s$), are negligible on the time scale of interest, $\kappa \gg \bar{g}_N \gg \gamma_j^p, \gamma_j^s $.

As the resonator field decays on a time scale much faster than that of the qubits we can adiabatically eliminate the field degrees of freedom \cite{Jonas}.  The master equation for the reduced density operator describing the qubits only is given by,
\begin{eqnarray}
\dot{\rho}_q = -i [\frac{\Delta_q}{2}J_z -\Delta_r \frac{ \bar{g}_N^2 }{|\Gamma|^2 } J_+J_-,\rho_q ] + \frac{\kappa}{2} \frac{ \bar{g}_N^2 }{ |\Gamma|^2}  \mathcal{D}[J_-] \rho_q,
\label{Adelimrho}
\end{eqnarray}
where, $\Gamma = \kappa/2 + i \Delta_r$, and it is assumed that the resonator is initially empty. The collective operators are defined as, $J_i = \sum^{N}_{j=1} \sigma^i_j$, where $i=\{ +,-,z\}$, and $\sigma^i_j$ denote the individual qubit Pauli matrices. Equation (\ref{Adelimrho}) corresponds to the superradiance master equation (SRME) that was derived for atomic ensembles, after a suitable parameter substitution \cite{Agarwal1974,NVSper, SR_PRLWoggon}.

We now seek an expression for the intensity of photons escaping the TLR, $I_{N} (t)$. Expanding  (\ref{Adelimrho}) in the Dicke basis \cite{Dicke}, the probability that the system is in one of the Dicke states $|l,m\rangle$ is, $P(l,m,t) =  \langle l,m | \rho_q (t) |l,m \rangle$, where, $ \mathbf{J}^2 |l, m\rangle = l(l+1) | l, m\rangle$, $J_z | l, m\rangle = 2m |  l, m\rangle$ and $N/2\ge l \ge |m| \ge 0  $. Using the SRME (\ref{Adelimrho}) the population rates follow \cite{Lee1},
\begin{eqnarray}\label{DifferenceEqn}
\dot{P}(l,m,t) &=& \frac{ \kappa \bar{g}_N^2 }{ |\Gamma|^2} [ (l-m)(l+m+1) P(l,m+1,t) \nonumber \\
&-& (l+m)(l-m+1) P(l,m,t)  ].
\end{eqnarray}
Here, we consider the initial condition that all qubits are excited, i.e., $P(l,m,0)=P(N/2,N/2, 0)$, although this approach can easily be extended to take into account other initial conditions \cite{Lee2}. As  $l$ is conserved by the SRME we introduce the variable, $n=l-m= 0, 1, ..., N$, which corresponds to the number of photons emitted from the resonator when the system is in the initial state $m=N/2$. Equation (\ref{DifferenceEqn}) can now be rewritten,
\begin{eqnarray}\label{DifferenceEqn2}
\dot{P}(n,\tau) &=&  (N-n+1) n P(n-1,\tau) - (N-n)(n+1) P(n,\tau).
\end{eqnarray}
where we have rescaled the time $\tau= \gamma t $ and $\gamma = \kappa \bar{g}_N^2 / |\Gamma|^2$. To proceed, we Laplace transform (\ref{DifferenceEqn2}), subject to the full excitation initial condition, $P(n,0)=\delta_{n,0}$,
\begin{eqnarray}\label{Lap11}
s Q(n,s) - \delta_{n,0} &=& (N-n + 1) n Q(n-1,s) \nonumber \\
&-& (N-n)(n+1) Q(n,s),
\end{eqnarray}
where, $Q(n,s)$ is the Laplace transform of $P(n, \tau)$. From  (\ref{Lap11}) we find $Q(0, s) = 1/(s+N)$. Continuing recursively we find,
\begin{equation}\label{Lap2}
Q(n>0,s) = \frac{1}{s+N} \prod^{n}_{i=1} \frac{(N-i+1)i}{s+ (N-i)(i+1)}.
\end{equation}
Upon inverting the transform (\ref{Lap2})  we obtain  the populations of the Dicke states, $P(n,\tau)$  \cite{Lee1}. The intensity of photons emitted from the TLR can be found from the Dicke state populations using, $I_N(\tau) = \frac{\partial}{\partial \tau} \sum_{n} n P(n,\tau)$. The system is superradiant if the maximum intensity, $I^{max}_{N}$, is greater than $N,$  the initial intensity of $N$ independent qubits \cite{NVSper}.

\begin{figure}
\centering
\subfigure[]{
{\includegraphics[scale=0.15]{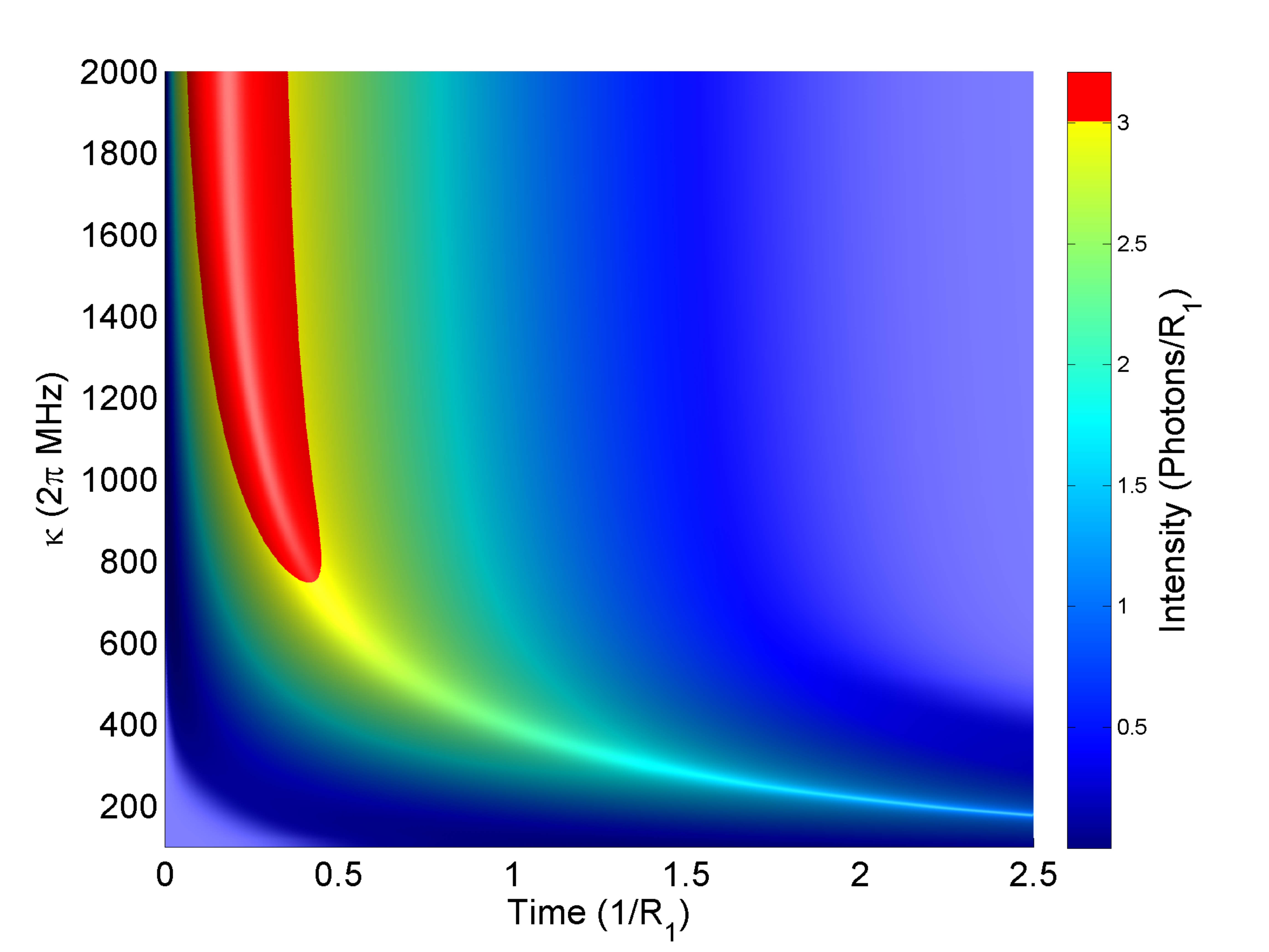}}
}
\subfigure[]{
{\includegraphics[scale=0.15]{SR3_NJPreply.pdf}}
}
\subfigure[]{
{\includegraphics[scale=0.15]{SR3_NJPreply.pdf}}
}
\caption[]{ (Color Online) Intensity of radiation from the resonator $I_N (R_1 t)$ obtained from numerical solution of the complete master equation, where $R_1 \equiv \Gamma(\Delta_r =0)= 4 \bar{g}_N^2  /\kappa$. (a), $N=3$. (b), $N=4$. (c), $N=5$.  Superradiant emission is colored red.  Parameters used are $g_{3,j} /2 \pi = (83.7, 85.7, 85.1)$ MHz \cite{Fink}, $g_{4,j} /2 \pi =( 69.4, 69.1, 68.6, 69.7)$ MHz,   $g_{5,j} /2 \pi = (59.0, 59.4, 59.9, 60.9, 60.7)$ MHz, and $(\kappa ,  \gamma_j^s , \gamma_j^p)/2 \pi = (2000, 0.19, 0)$ MHz. \label{IntensityBurst} }
\end{figure}

Solving the SRME for the intensity profiles for experimentally relevant systems of  TLRs with $N=3,4$ or $5$ qubits  we find,
\begin{subequations}
 \label{Intensity}
\begin{eqnarray}
I_3 (\tau) &=& 3e^{-3 \tau} (12 \tau-7 )+ 24 e^{-4 \tau}. \label{Intensity3} \\
I_4 (\tau) &=&  (72 \tau+96)e^{-6 \tau}+4 e^{-4 \tau} (36 \tau -23), \label{N45}\\
I_5 (\tau) &=& \frac{5}{3} [162e^{-9 \tau} +16 e^{-8 \tau} (24 \tau -1)  + e^{-5 \tau} (240 \tau-143)].\label{N452}
\end{eqnarray}
\end{subequations}
The system is superradiant if the maximum intensity, $I^{max}_{N}$, is greater than $N$,  the initial intensity of $N$ independent qubits \cite{NVSper}. As  $I^{max}_{3}\approx 3.2  > 3$, $I^{max}_{4} \approx 4.9 > 4$, and, $I^{max}_{5} \approx 6.9 > 5$, these systems exhibit superradiance.

In Fig.~\ref{IntensityBurst} the intensity of emission from the ensembles is shown for a range of  experimentally feasible parameters. It is clear that superradiance (colored red) can be observed over a large range of cavity decay rates, $\kappa$. The approximate solutions (\ref{Intensity3}-c) closely resemble the results in Fig.~\ref{IntensityBurst}  for  $\kappa \gg \bar{g}_N \gg \gamma_j^s, \gamma_j^p$. For each system the superradiant peak is large enough to be easily resolved using existing detection schemes \cite{WallraffCorrelation, MB, MC}.

The above results demonstrate that it is possible to observe small sample superradiance in a TLR in the presence of energy relaxation and non-uniform coupling strengths. Superradiance can also be observed for $N>5$ qubits, provided $\kappa \gg \bar{g}_N \gg \gamma_j^s, \gamma_p^s$ and $g_{N,j} \approx \bar{g}_N$, however this may be more difficult to achieve experimentally. For a TLR with resonance frequency $\omega_r$, the length determines the number of qubits that can be effectively coupled at the maximal coupling strength as each qubit is typically placed at the field antinodes. When $N$ is increased the coupling rate to each qubit is reduced as the coupling rate $\bar{g}_N$ is inversely related to the mode volume \cite{wiringup}. As $\bar{g}_N \rightarrow \gamma^s_j$, losses due to energy relaxation and coupling to subradiant states will dominate the dynamics and superradiant effects will become more difficult to observe. However as the intensity scales as, $I^{max}_{N}\propto N^2$, the resulting dynamics will be the determined from the competition between the superradiant intensity enhancement and the aforementioned losses.

The proposal presented here is superior to existing superradiance experiments in three key aspects: initial state preparation, measurement and losses. In circuit-QED arbitrary initial states can be prepared at high fidelity using a sequence of quantum logic gates \cite{Dicarlo3Qubit}, whereas, in bulk solids and atomic gases a pulse is fired into the medium, randomly exciting sections of the ensemble \cite{SR_QDs, SRBEC2003,Clearest, Lamb, HarocheSR}. The lack of control over the initial state in the latter schemes severely limits the possible initial states that can be prepared and can lead to light being re-absorbed by the atomic system resulting in delays and distortions of the exiting superradiant pulse, making analysis extremely complicated. Recent advances in circuit-QED have also lead to high fidelity, single-shot joint multi-qubit measurements \cite{Dicarlo3Qubit}, which can be used to completely characterise the state of the ensemble during emission. This can also be used to monitor the build up of field mediated correlations between individual qubits. Such quantum correlations are a hallmark of the superradiant process \cite{Agarwal1974, HarocheReview}. In atomic demonstrations of superradiance, characterisation of the quantum states of the atomic system using techniques such as quantum state tomography has not been possible. Instead one monitored the emission from the ensemble to characterize superradiance. However in the superconducting case, full quantum state tomography can be performed on the superconducting qubits and detailed superradiance mediated correlations can be fully mapped out. Lastly, circuit-QED systems have relatively uniform qubit-field coupling rates and remarkably low rates of decoherence and energy relaxation. In other implementations, atomic collisions combined with spatially varying qubit-field coupling rates lead to losses which can only be overcome by using large ensembles. By using the low loss circuit-QED system presented here, it is no longer necessary to use such large ensembles to observe superradiance. Small sample superradiance may have applications in quantum information, including decoherence free subspace state encoding using subradiant states \cite{KnightDFS} and superradiant state readout \cite{Briggs}.

\subsection{Discussion of the two level approximation to the transmon \label{ThreeLevel}}


In this paper we make the two level approximation to the transmon qubit, whereas infact the transmon is an anharmonic oscillator \cite{transmonpaper}. With several excitations in the system it may be possible to induce transitions to higher energy levels than the qubit levels, slowing the superradiant decay. Here, we analyse the effect that a third level for each transmon has on the dynamics. Firstly, we note that the two level approximation is a very common approximation in the literature \cite{transmonpaper, DynamicalBistability, TransmonTheoryGatePaper} and has been verified in several experiments \cite{Fink, TwoLTransmonExp,LifeAfterChargeNoise}. Secondly, the transmon's anharmonicity can be varied considerably by altering the ratio of the Josephson energy to the charging energy ($E_J/E_C$). From \cite{transmonpaper} it can be seen that the anharmonicity, $\alpha_r$, can be as large as $11\% $ of the qubit level's transition frequency for $E_J/E_C \sim 17.5$ (i.e $\alpha_r/ 2 \pi \sim 660$ MHz at a qubit transition frequency of $\omega_r/2 \pi =6$GHz). This can be achieved with a modest increase in the transmons susceptibility to charge noise. At this level of anharmonicity, the third level is far detuned from the transition frequency and has very little effect on the system dynamics.

\begin{figure}
\centering
{\includegraphics[scale=0.45]{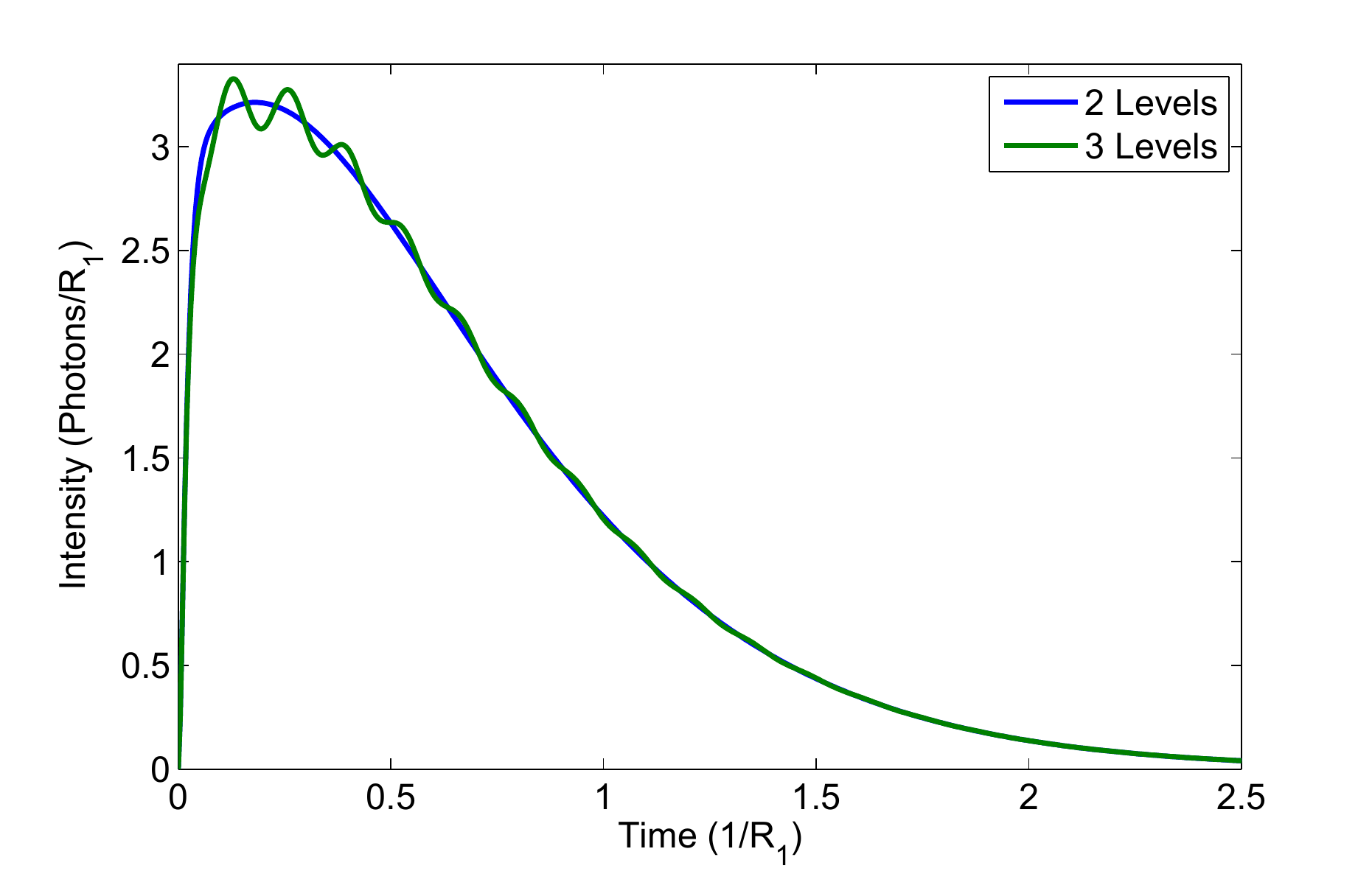}}
\caption{  Intensity of radiation from the resonator $I_3 (R_1 t)$ obtained from numerical solution of the complete master equation with the two level and three level approximations for each transmon.  Parameters used are $g_{3,j} /2 \pi = (83.7, 85.7, 85.1)$ MHz \cite{Fink} and $(\omega_r, \alpha_r, \kappa ,  \gamma_j^s , \gamma_j^p)/2 \pi = (6000, 660, 2000, 0.19, 0)$ MHz. The coupling rate to the third transmon level, $G_{3,j}$, is approximated using $G_{3,j} = \sqrt{2} g_{3,j}$ \cite{{Sqrt2}}.
 }
 \label{TwoLevelsVsThreeLevels}
\end{figure}

In the bad cavity limit the photons escape immediately after the transmon decays, therefore there is no possibility for the decay of one transmon to excite higher levels of another transmon. In our simulations, $\kappa$ is much larger than $\bar{g}_N$ to ensure this limit is valid. However, as $\kappa$ is finite, there is a small probability of the decay of one transmon exciting another transmon to a higher level. To see the effect, third level transmons in our simulations. We find that the population of the third level is negligible. However, we do find the third level induces small oscillations in the intensity from the cavity (Fig.~\ref{TwoLevelsVsThreeLevels}). This is a well known effect \cite{StojanCorrelations, 2011PRA} whereby a far detuned three level system acts like a two level system with minor oscillations at a frequency proportional to the detuning of the third level.


\section{Phase Multistability \label{PM}}
We now study another collective effect that can be observed in the same system - phase multistability. Phase multistability is the phenomenon where a coherently driven multi-qubit cavity system displaces the cavity field in phase depending on the collective state of the qubits. To observe phase multistability we return to the $g>\kappa$ regime in Eq. (\ref{ME_NCPB}) and resonantly drive the system. The system is now described by the following master equation,
\begin{eqnarray} \label{ME_PM}
\dot{\rho} &=& -i [  \sum^N_{j=1} g_{N,j} ( \sigma^-_j a^{\dagger} + a \sigma^+_j), \rho] + [  \mathcal{E} ( a^{\dagger} -  a ) , \rho]  \\ \nonumber
&+& \frac{\kappa}{2}  \mathcal{D}[a] \rho + \sum^N_{j=1} \frac{\gamma^s_j}{2} \mathcal{D}[ \sigma_j^-] \rho
\end{eqnarray}
where, we assume the drive field and the qubits are resonant with the TLR ($\Delta_r, \Delta_{q, j}=0$). Furthermore, we note that dephasing is negligible for transmon qubits, $\gamma^p_j\approx 0$.

Single qubit phase bistability has proven very difficult to experimentally achieve as the qubit-cavity system needs to fulfill the requirements of the strong coupling regime, $g > \kappa , \gamma^s $. To date only a single experiment has detected signatures of phase bistability \cite{Armen}, however, the modest ratio of $g/\kappa$ and various other experimental difficulties meant that the results were only marginally conclusive.  Phase multistability for $N>1$ has yet to be observed. In this section we show that the circuit-QED system presented here, with the addition of coherent driving can display phase multistability and should present very clear experimental signatures.

\subsection{Phase Bistability}
Phase bistability can be described in the $ \gamma^s \rightarrow 0$ limit as follows: at large intra-cavity photon numbers, $n$, the difference in energy between successive Jaynes-Cummings  manifolds $|n, \pm \rangle \leftrightarrow |n+1, \pm \rangle$ is $ E^{\pm}_{n+1}-E^{\pm}_{n} \approx \omega_r \pm g/2\sqrt{n}$, where $\omega_r$ is the frequency of the resonator and  $|n, \pm \rangle = 1/\sqrt{2} ( |n+1,g \rangle  \pm   |n,e\rangle)$. Strong driving of the TLR at $\omega_r$ will populate large $n$ Jaynes Cummings eigenstates and cause transitions along two separate pathways, $|n, \pm \rangle \leftrightarrow |n+1, \pm \rangle$. This driving is detuned from the transition frequency of each path by the manifold dependent detuning, $\pm g/(2\sqrt{n})$. As the field is displaced in phase depending on the detuning, the field acquires a phase depending on the state of the qubit. The steady state density matrix of the system ($\rho_{ss}$) can be approximated by a mixture of two uncoupled, damped harmonic oscillators. These oscillators are driven off-resonantly and displaced in phase due to the different detunings of the effective driving fields experienced by each oscillator  \cite{KillinSingle,Alsing}. As $t\rightarrow \infty$ the coherent state amplitude of each oscillator $| \alpha_{ss}^{\pm} \rangle$ is $\alpha_{ss}^{\pm} = f(2 \mathcal{E} f \pm  i g )/\kappa$, where $f = \sqrt{ 1-(g/2\mathcal{E})^2 }$. The two coherent states $| \alpha_{ss}^{\pm} \rangle$ can be detected by homodyne measurement of the resonator field \cite{MD,ME,MF}.

Now relaxing the assumption, $ \gamma^s \rightarrow 0$, we consider the case where, $f\rightarrow 1$, and in the steady state the system is in $| \alpha_{ss}^{+}, +\rangle$.  The system will remain in this state until energy relaxation causes the qubit transition $|+\rangle \xrightarrow{\gamma^s} 1/\sqrt{2} ( | + \rangle + | -\rangle)$. After a short transient regime, the system will either reach the steady state,  $| \alpha_{ss}^{+}, +\rangle$ or $| \alpha_{ss}^{-}, -\rangle$ with equal probability. Due to this effect, in a single quantum trajectory the imaginary part of the cavity amplitude randomly switches between $\pm g /\kappa$ at the rate $\gamma^s/4$  \cite{Alsing}.

To observe phase bistability experimentally, it is necessary to resolve the two coherent states $|\alpha_{ss}^{\pm}\rangle$. For strong driving this requires the ratio $g/ \kappa$ to be as large as possible. Also, as energy relaxation couples the two transition pathways, it can be shown \cite{KilinMulti} that phase bistability only occurs when  $\gamma^s < 2\kappa$. As circuit-QED systems can fulfil each of these requirements, and does not suffer from problems associated with moving atoms in cavity QED \cite{Armen}, circuit-QED is  a preferable system for the observation of phase bistability.

\begin{figure}[t]
\centering
{\includegraphics[scale=0.56]{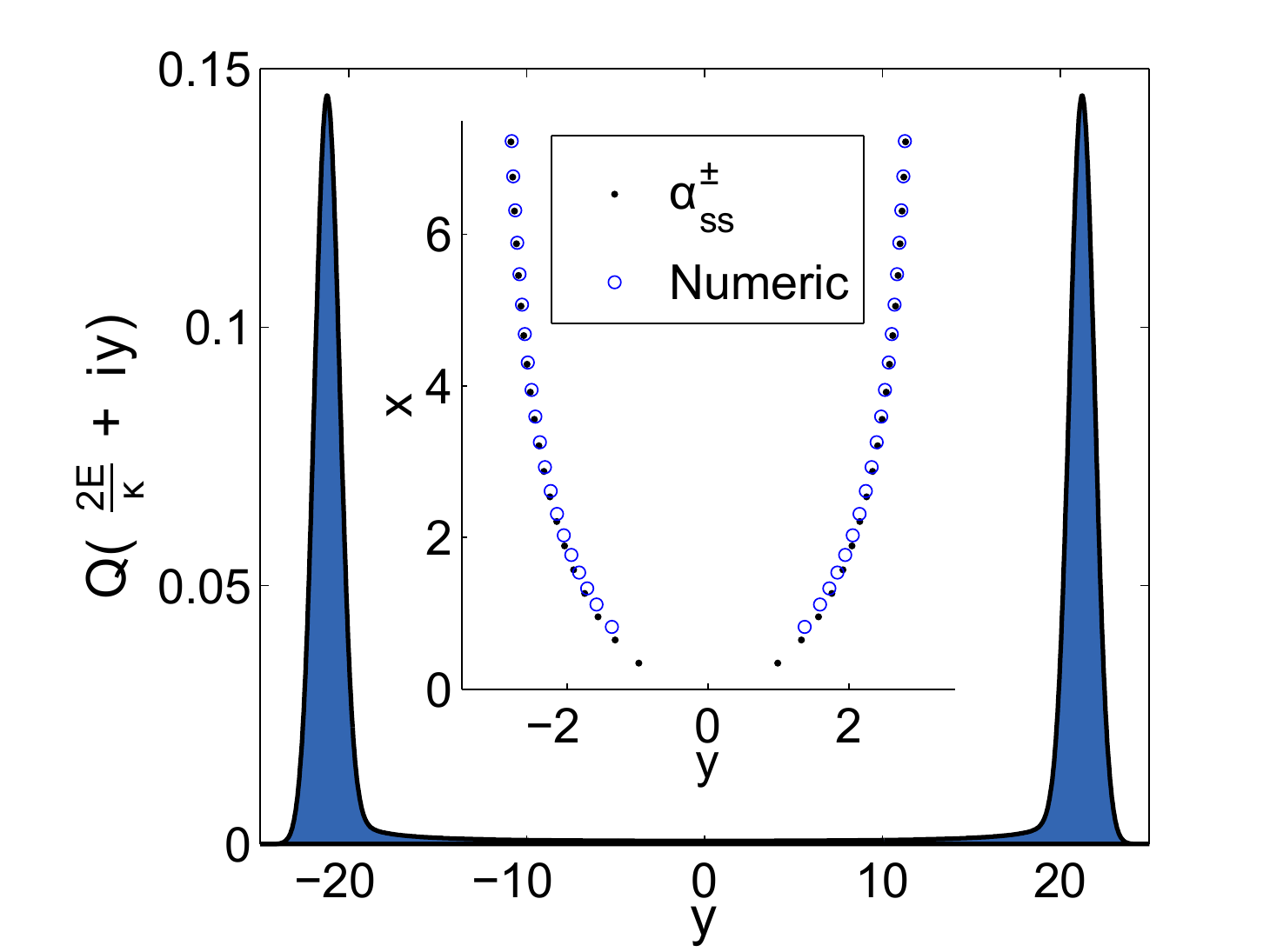}}
\caption{
The  $Q$-function in Eq. (\ref{AnalyticQ}) at $x= 2 \mathcal{E}/\kappa$ with parameters $(g, \kappa,  \gamma^s)/2 \pi = (85, 4, 0.19)$ MHz. \textit{Inset:}  The location of the two peaks in the steady state $Q$-function for a range of  driving strengths $\mathcal{E}/ 2\pi = 45-117.6$ MHz, increasing from bottom to top.  The following parameters were used: $(g, \kappa ,  \gamma^s)/2 \pi = (85, 28.3, 0.19)$ MHz.
}

 \label{SAQ}
\end{figure}

For large driving amplitude the steady state $Q$-distribution can be found after appropriate transformations of the density matrix \cite{KillinSingle},
\begin{equation}\label{AnalyticQ}
 Q(x+ iy) = \frac{2e^{-\left(x-\frac{2\mathcal{E}}{\kappa}\right)^2}}{ 2^{\frac{\gamma^s }{\kappa}} \pi  \beta(\frac{\gamma^s }{2 \kappa},\frac{\gamma^s }{2 \kappa})} \int _{-1}^{1}  \frac{e^{-(\frac{g}{ \kappa}z-y)^2}}{(1- z^2)^{(1-\frac{\gamma^s }{2 \kappa})}} dz,
\end{equation}
where, $Q(\alpha) = \langle \alpha | \rho_{ss} | \alpha \rangle / \pi, \alpha = x+iy$ and $\beta(a,b)$ is the beta function.   A cross section of this function is shown in Fig.~\ref{SAQ} for realistic circuit-QED parameters. It is clear that  $|\alpha_{ss}^{\pm}\rangle$ can be resolved using homodyne measurement of the field \cite{WallraffCorrelation, Armen}. In contrast to \cite{Armen}, we see that a circuit-QED demonstration of this effect will be dramatic and unambiguous  demonstration of phase bistability.

The inset in Fig.~\ref{SAQ} compares the peak locations $\alpha_{ss}^{\pm}$ to those obtained from numerical solution of the steady state density matrix for a range of driving amplitudes. At small driving amplitudes the system is too anharmonic to approximate as two harmonic oscillators. However, for larger amplitudes the two solutions coincide. Phase bistability has been the basis for several proposals including ultralow energy optical switching \cite{Armen}, quantum feedback and qubit measurement \cite{Mabuchi}.

Similar to section \ref{ThreeLevel} we review the validity of the two level approximation to the transmon. The higher levels will lead to more transition pathways and therefore more peaks in the steady state $Q$-distribution. However, if the transmon is strongly anharmonic, these transitions will be suppressed and the additional peaks will be very minor. As strong driving can populate higher lying states, to ensure the validity of the two level approximation the drive frequency needs to be significantly less than the anharmonicity, $\alpha_r \gg \mathcal{E} > g/2 > \kappa > \gamma^s/2$. Furthermore, the populations of the higher levels can be reduced by enhancing the rate of energy relaxation $\gamma^s$.

\subsection{Phase Multistability}
When there are $N$ qubits in the resonator an analogous  phenomenon occurs: phase multistability \cite{KilinMulti}. Similarly to the single qubit case, phase multistability results from the energy structure of Tavis-Cummings manifolds at large excitation. For a given $l$, there are, $2 l+1$, transitions between the Dicke states $|n\rangle | l,m\rangle \leftrightarrow |n+1\rangle | l,m\rangle$ \cite{Dicke}. The difference in energy between successive Tavis-Cummings manifolds at large intra-cavity photon number is, $ E^{(l,m)}_{n+1}-E^{(l,m)}_{n} \approx \omega_r + m \bar{g}_N /\sqrt{n}$  \cite{KilinMulti}.  Proceeding as before, assuming $\gamma^s_j \rightarrow 0$ and $g_{N,j} \approx \bar{g}_N$,  strong resonant driving of the TLR leads to transitions on $2 l+1$ separate ladders. As the field is displaced in phase depending on the detuning, the field acquires a phase depending on the state of the qubit. The steady state density matrix of the system can be approximated by a mixture of $2 l+1$ damped, driven uncoupled harmonic oscillators. Each harmonic oscillator has the coherent state amplitude,
\begin{equation}\label{Nqfunc}
\alpha_{ss}^{(l,m)} = 2 f_m ( \mathcal{E} f_m  + i m \bar{g}_N )/\kappa,
\end{equation}
where, $f_m =  \sqrt{1- (m \bar{g}_N/\mathcal{E} )^2}$.

\begin{figure}[t]
\begin{center}
 {\includegraphics[scale=0.62]{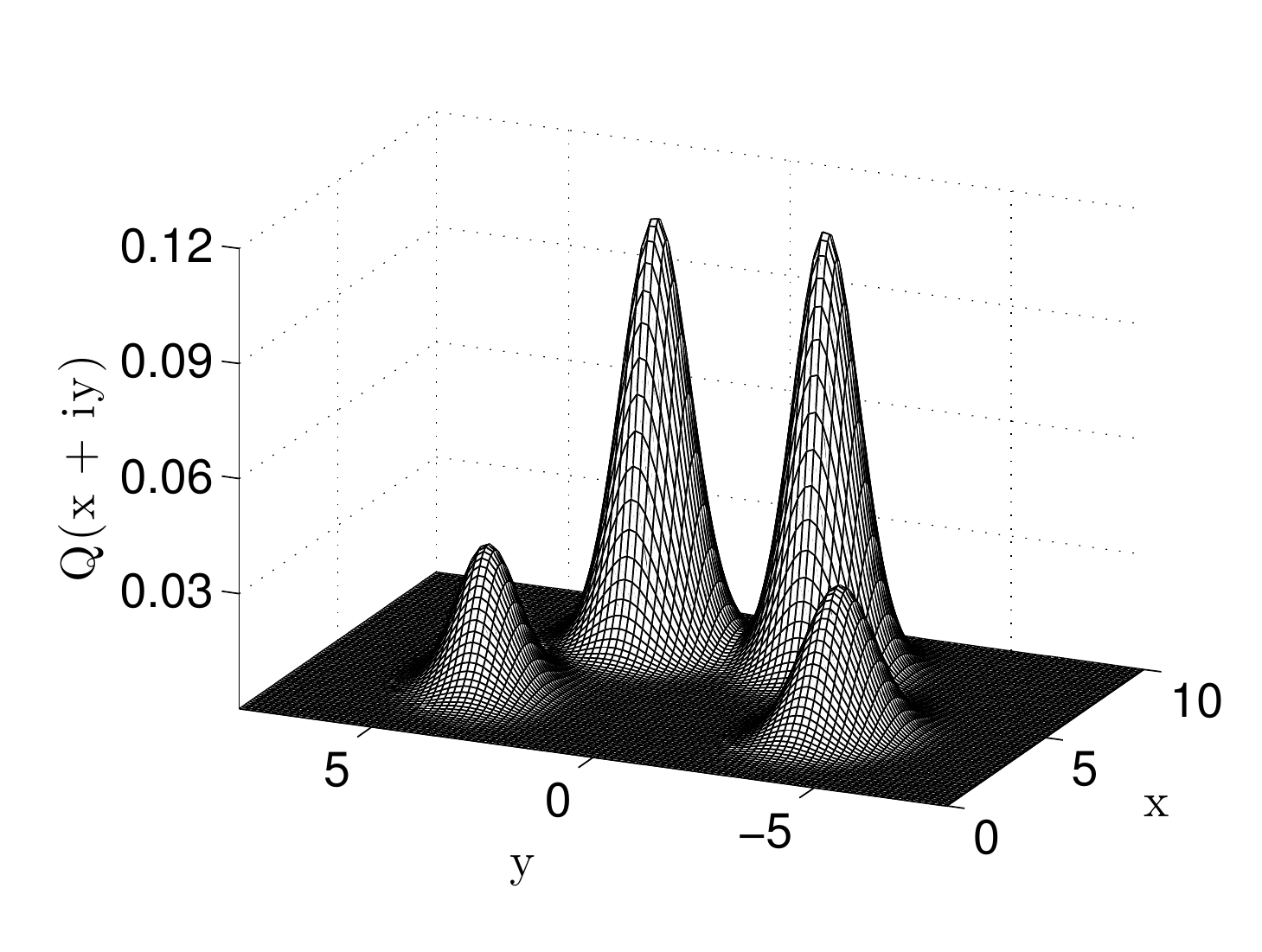}}
\caption{ Steady state $Q$-function for the field of a TLR with three qubits obtained from  numerical solution of equations (\ref{ME_NCPB}) with $H_{\mathcal{E}}$. Center peaks correspond to $\alpha_{ss}^{(3/2,\pm 1/2)}$, whilst the other peaks correspond to $\alpha_{ss}^{(3/2,\pm 3/2)}$ (see Eq. (\ref{Nqfunc})). Parameters used are $(\mathcal{E}, \kappa,  \gamma_j^s ) /2 \pi= (169.6, 42.4, 0.19)$ MHz and $g_{3,j} /2 \pi = (83.7, 85.7, 85.1)$ MHz \cite{Fink}. Note that the distance between the peaks can be increased by choosing smaller $\kappa$.
}
\end{center}
 \label{ThreeAtomQ}
\end{figure}

Due to the small ratio of $\bar{g}_N / \kappa$ and the inability to couple several qubits identically to a common field mode, phase multistability has not been experimentally demonstrated. However, using existing circuit QED parameters Fig.~\ref{ThreeAtomQ} demonstrates that phase multistability can be observed. The four peaks in the steady state $Q$-distribution in Fig.~\ref{ThreeAtomQ} are the four field displacements in phase space corresponding to the collective qubit states with $l=-3/2,-1/2, 1/2, 3/2$. The peak positions $\alpha_{ss}^{(l,m)}$ coincide with the numerical solution to high precision. The centre two peaks are larger because the $m=1/2$ transitions are driven closer to resonance than the  $m=3/2$ transitions. Phase multistability may have similar applications as phase bistability, i.e in quantum feedback \cite{Mabuchi} and ultralow energy microwave switching \cite{Armen, MG, MH}. Moreover, it may be possible to use phase multistability to perform a collective measurement of the system to determine the collective spin, $m$ \cite{Mabuchi}.

\section{Conclusion\label{con}}
In summary, we have demonstrated that two previously unobservable collective quantum optical phenomena,  small sample superradiance and phase multistability, can be observed in a circuit-QED based system using existing experimental parameters. The proposal presented here should provide a definitive and unambiguous demonstration of superradiance, with the added benefit of full quantum state tomography on the qubits. Owing to its the experimental tractability we expect that this approach will be indispensable for a range experiments on collective phenomena.

\section{Acknowledgments}
We thank L. Willink, G. J. Milburn and G. K. Brennen for useful discussions. This work was
supported by ARC (Centre for Quantum Computing Technology, DP0986932 and DP1094758).


\end{document}